\documentclass[]{aa}  
\usepackage{graphicx}
\usepackage{amsmath}
\usepackage{txfonts}
\usepackage{fourier}
\usepackage{natbib}
\usepackage{mathrsfs}

\begin{document}

\newcommand{\teff}{T$_{\rm eff}$}
\newcommand{\logg}{$\log${(g)}}
\newcommand{\met}{$[$Fe/H$]$}

%
   \title{The GAPS Programme at TNG
   \thanks{Based on observations made with the Italian Telescopio Nazionale Galileo (TNG) operated by the Fundación Galileo Galilei (FGG) of the Istituto Nazionale di Astrofisica (INAF) at the Observatorio del Roque de los Muchachos (La Palma, Canary Islands, Spain).}\ 
   \thanks{Table~\ref{tab:rvs} is only available in electronic form at the CDS via anonymous ftp to cdsarc.u-strasbg.fr (130.79.128.5) or via http://cdsweb.u-strasbg.fr/cgi-bin/qcat?J/A+A/}\ 
   \thanks{This is the Accepted Manuscript version of an article accepted for publication in Astronoym \& Astrophysics. IOP Publishing Ltd is not responsible for any errors or omissions in this version of the manuscript or any version derived from it. The Version of Record is available online at the Journal's website.}}

   \subtitle{XXIX. No detection of reflected light from 51~Peg~b using optical high-resolution spectroscopy}

   \author{G. Scandariato\inst{1}{\thanks{e-mail: gaetano.scandariato@inaf.it}}, F. Borsa\inst{2}, D. Sicilia\inst{1,3}, L. Malavolta\inst{1,3}, K. Biazzo\inst{4}, A. S. Bonomo\inst{5}, G. Bruno\inst{1}, R. Claudi\inst{6}, E. Covino\inst{7}, P. Di Marcantonio\inst{8}, M. Esposito\inst{9}, G. Frustagli\inst{2,10},  A.F. Lanza\inst{1},  J. Maldonado\inst{11}, A. Maggio\inst{11}, L. Mancini\inst{5,12,13}, G. Micela\inst{11}, D. Nardiello\inst{6,14}, M. Rainer\inst{15}, V. Singh\inst{1}, A. Sozzetti\inst{5}, L. Affer\inst{11}, S. Benatti\inst{11}, A. Bignamini\inst{8}, V. Biliotti\inst{15}, R. Capuzzo-Dolcetta\inst{16}, I. Carleo\inst{17}, R. Cosentino\inst{18}, M. Damasso\inst{5}, S. Desidera\inst{6}, A. Garcia de Gurtubai\inst{18}, A. Ghedina\inst{18}, P. Giacobbe\inst{5}, E. Giani\inst{15}, A. Harutyunyan\inst{18}, N. Hernandez\inst{18}, M. Hernandez Diaz\inst{18}, C. Knapic\inst{8}, G. Leto\inst{1}, A. F. Mart\'inez Fiorenzano\inst{18}, E. Molinari\inst{19}, V. Nascimbeni\inst{6}, I. Pagano\inst{1}, M. Pedani\inst{18}, G. Piotto\inst{3}, E. Poretti\inst{18}, H. Stoev\inst{18}
      }

   \institute{
INAF -- Osservatorio Astrofisico di Catania, Via S.Sofia 78, I-95123, Catania, Italy
    \and
INAF -- Osservatorio Astronomico di Brera, Via E. Bianchi 46, 23807 Merate, Italy
    \and
Dipartimento di Fisica e Astronomia Galileo Galilei, Universit\`a di Padova, Vicolo dell'Osservatorio 3, I-35122, Padova, Italy
    \and
INAF -- Osservatorio Astronomico di Roma, Via Frascati 33, I-00040, Monte Porzio Catone (RM), Italy
    \and
INAF -- Osservatorio Astrofisico di Torino, Via Osservatorio 20, I-10025, Pino Torinese, Italy
    \and
INAF -- Osservatorio Astronomico di Padova, Vicolo dell’Osservatorio 5, I-35122, Padova, Italy
    \and
INAF -- Osservatorio Astronomico di Capodimonte, Salita Moiariello 16, 80131 Napoli, Italy
    \and
INAF -- Osservatorio Astronomico di Trieste, Via Tiepolo 11, 34143 Trieste, Italy
    \and
Th\"uringer Landessternwarte Tautenburg, Sternwarte 5, 07778, Tautenburg, Germany
    \and
Dipartimento di Fisica G. Occhialini, Universit\`a degli Studi di Milano-Bicocca, Piazza della Scienza 3, 20126 Milano, Italy
    \and
INAF -- Osservatorio Astronomico di Palermo, Piazza del Parlamento, 1, I-90134 Palermo, Italy
    \and
Department of Physics, University of Rome \lq\lq Tor Vergata\rq\rq, Via della Ricerca Scientifica 1, I-00133, Rome, Italy 
    \and
Max Planck Institute for Astronomy, Königstuhl 17, D-69117, Heidelberg, Germany 
    \and
Aix Marseille Univ, CNRS, CNES, LAM, Marseille, France 
    \and
INAF -- Osservatorio Astrofisico di Arcetri, Largo Enrico Fermi 5, I-50125 Firenze, Italy
    \and
Dip. di Fisica, Sapienza, Università di Roma,  Piazzale Aldo Moro, 2, I-00185, Roma, Italy
    \and
Astronomy Department and Van Vleck Observatory, Wesleyan University, Middletown, CT 06459, USA
    \and
Fundaci\'on Galileo Galilei - INAF, Rambla Jos\'e Ana Fernandez P\'erez 7, E-38712, Bre\~na Baja, TF - Spain
    \and
INAF -- Osservatorio Astronomico di Cagliari \& REM, Via della Scienza, 5, I-09047 Selargius CA, Italy
   }


  \abstract{The analysis of exoplanetary atmospheres by means of high-resolution spectroscopy is an expanding research field which provides information on chemical composition, thermal structure, atmospheric dynamics and orbital velocity of exoplanets.}
  {In this work, we aim at the detection of the light reflected by the exoplanet 51~Peg~b employing optical high-resolution spectroscopy.}
  {To detect the light reflected by the planetary dayside we use optical HARPS and HARPS-N spectra taken near the superior conjunction of the planet, when the flux contrast between the planet and the star is maximum. To search for the weak planetary signal, we cross-correlate the observed spectra with a high S/N stellar spectrum.}
  {We homogeneously analyze the available datasets and derive a $10^{-5}$ upper limit on the planet--to--star flux contrast in the optical.}
  {The upper limit on the planet--to--star flux contrast of $10^{-5}$ translates into a low albedo of the planetary atmosphere ($\rm A_g\lesssim0.05-0.15$ for an assumed planetary radius in the range $\rm 1.5-0.9~R_{Jup}$, as estimated from the planet's mass).}


\authorrunning{G. Scandariato et al.}

\titlerunning{The GAPS Programme: No detection of reflected light from 51~Peg~b}

   \maketitle
%

\section{Introduction}

The atmospheric characterization of known exoplanets has tremendously developed since the first detection of an exoplanet atmosphere \citep{Charbonneau2002}.

Transiting exoplanets are the most favourable targets for atmospheric characterizations. During transit, the outer layers of the gaseous envelopes of the planet will filter the background stellar light and imprint features due to diffuse scattering and line absorption. In this regard, spectroscopic and photometric observations have proven to be a powerful tool for the atmospheric study of these bodies, both from space \citep[e.g.][]{Vidal2004,Sing2009,Sotzen2020,Garhart2020} and from the ground \citep[e.g.][]{Nascimbeni2015,Mancini2017,Vissapragada2020,Guilluy2020,Sicilia2020}. Moreover, with the improvement of the available instrumentation, it has been possible to detect and analyze the phase curves and secondary eclipses of exoplanets, leading to the characterization of the planetary dayside \citep{Stevenson2014,Parmentier2018,Kreidberg2018,Singh2020}.

Despite the lack of information from transits and eclipses, additional investigations have also been directed to non-transiting exoplanets with particularly interesting properties. In this respect, near-infrared and optical spectroscopy have been successfully adopted: in particular, the former aims at investigating the emitted spectrum of the planetary dayside \citep{Brogi2012,Birkby2017}, while the latter allows for the examination of the stellar reflected spectrum \citep{Martins2015}. Both techniques constrain the chemical composition and thermal structure of the planetary atmosphere. Moreover, if phase-resolved high-resolution spectroscopy is available, then it is also possible to measure the planet’s orbital velocity. This information is particularly valuable as it leads to determine the inclination of the orbital plane and, by consequence, the true mass of the non-transiting planet. Thanks to a technique developed for double-lined spectroscopic binaries \citep{Hilditch2001}, it is indeed possible to break the degeneracy between the planetary mass and the inclination of the orbital plane.

51~Peg~b (HD~217014~b) is the first exoplanet discovered around a solar-type star using the radial velocity technique \citep{Mayor1995}. So far, the search of planetary transits has failed \citep{Mayoretal1995,Walker2006}, and photometric techniques cannot provide the orbital inclination and, consequently, the mass of the planet. Hence, the investigation of the planetary spectrum, either reflected or emitted, has been motivated by two reasons: the characterization of the planetary atmosphere and the measurement of the inclination of its orbit. The first successful high-resolution near-infrared spectroscopic analysis of 51~Peg~b has been reported by \citet{Brogi2013} and later corroborated by \citet{Birkby2017}, while the optical spectrum has been detected and analyzed by \citet{Martins2015} and \citet{Borra2018}. All these works suggest an orbital inclination between 70$^\circ$ and 80$^\circ$, and a planetary mass of approximately half a Jovian mass.

Nonetheless, the detection of the optical spectrum is still debated. Such detection would imply that 51~Peg~b has an unusually high geometric albedo for the class of hot Jupiters (HJs), and makes it stand out in the search for elusive correlations between atmospheric properties and stellar irradiation \citep{Heng2013}. Moreover, the same optical spectra have been reanalyzed by \citet{DiMarcantonio2019}, who do not reproduce the claimed signal, yet cautioning that their method could not reach the accuracy level needed for the detection of the signal.

In this perspective, we here reanalyze and extend the previous analysis of optical spectra using all the available data in the HARPS and HARPS-N archives. Firstly, in Sect.~\ref{sec:background} we provide a detailed mathematical framework, while in Sect.~\ref{sec:observations} we describe the analyzed datasets. In Sect.~\ref{sec:orbit} we refine the ephemeris of 51~Peg~b using the full set of radial velocity data available in the literature, complemented with the newest measurements. With our refined orbital solution, we chase the phase-resolved planetary signal as discussed in Sect.~\ref{sec:analysis}. Finally, in Sect.~\ref{sec:discussion} we draw our conclusions.

\section{Theoretical background}\label{sec:background}

The spectroscopic observation of a planet-host star returns a spectrum which is in principle the superposition of the stellar spectrum and the planetary spectrum, whether the latter is due to the reflected stellar light and/or planet thermal emission. In particular, for the case of 51~Peg~b, we expect that the contrast between the reflected flux and the stellar flux is of the order of $10^{-5}$--$10^{-4}$ in the optical domain (see below). As for the thermally emitted, we can assume a stellar effective temperature of 5790~K and a stellar radius of 1.20~$\rm R_\sun$ \citep{Fuhrmann1997}, while for the planet we can assume an approximate radius of 1~$\rm R_{Jup}$ (see Sect.~\ref{sec:discussion}) and a conservative dayside temperature of 2000~K \citep{Brogi2013,Birkby2017}. Under these hypothesis, integrating the black body intensities in the spectral range covered a typical echelle spectrograph (3000--7000~\AA), the thermal emission is an order of magnitude fainter than the expected reflected spectrum. In the following we will thus neglect the contribution of the thermal emitted spectrum.

To set up the theoretical background needed for the interpretation of our results we follow \citet{Perryman2018} and references therein.

\subsection{The planet spectrum}

If we define the planetary-to-stellar flux contrast at the orbital phase $\phi$ as $\epsilon(\phi)$, then the spectrum reflected by the planet is given by:
\begin{equation}
F_p=\epsilon\ F_\star\left(\lambda\left(1+\frac{v_p}{c}\right)\right),\label{eq:fp}
\end{equation}
where $v_p$ is the phase-dependent radial velocity of the planet in the stellar rest frame, $c$ is the speed of light in vacuum and $F_*(\lambda)$ is the stellar spectrum in the stellar rest frame. If, for the sake of simplicity, we assume that the flux contrast $\epsilon$ does not depend on $\lambda$, and that the star slowly rotates in the planet's rest frame, then the planetary spectrum is basically a rescaled version of the stellar spectrum Doppler-shifted by the radial velocity of the planet $v_p$ in the stellar rest frame.

Assuming that the planet is in a circular orbit, which is a reasonable approximation for 51~Peg~b (as we derive in Sect.~\ref{sec:orbit}), then we can write
\begin{equation}
v_p(\phi)=K_p\sin(2\pi\phi),\label{eq:vp}    
\end{equation}
where $K_p$ is the radial velocity amplitude of the planet referred to the stellar rest frame, while $\phi$ is the orbital phase ranging in the 0--1 interval ($\phi=0$ corresponds to the planetary inferior conjunction).

As for the contrast $\epsilon$ in Eq.~\ref{eq:fp}, we first define the phase angle $\alpha$ as the star--planet--observer angle given by:
\begin{equation}
\cos\alpha=-\sin i\ \cos(2\pi\phi),\label{eq:cosalpha}
\end{equation}
where $i$ is the orbital inclination. The phase angle $\alpha$ determines the phase function $g(\alpha)$, which models the amount of the light reflected towards the observer. In an edge-on orbit ($i=90^\circ$), the phase function is 0 during a transit (when only the night side of the planet is visible) and increases to 1 during a secondary eclipse, that is we would see the full day-side of the planet if it were not occulted by its host star. For the sake of simplicity, we will assume that the planet follows Lambert's scattering law, in which case the phase function is obtained analytically and is given by:
\begin{equation}
g(\alpha)=\frac{\sin\alpha+(\pi-\alpha)\cos\alpha}{\pi}.\label{eq:phaseFunction}
\end{equation}

The star--planet contrast thus depends on the orbital phase of the planet through the phase angle $\alpha$ as:
\begin{equation}
\epsilon(\alpha)=\epsilon_{max}g(\alpha),~{\rm with}~\epsilon_{max}=A_g\left[\frac{R_{\rm p}}{a}\right]^2,\label{eq:epsilon}
\end{equation}
where:
\begin{itemize}
    \item $[R_{\rm p}/a]^2$ is a scaling geometrical factor which sets the amount of stellar flux incident on the planet, and depends on the planetary radius $R_{\rm p}$ and the orbital semi-major axis $a$;
    \item $A_g$ is the geometric albedo of the planet;
    \item the phase function $g(\alpha)$ modulates the maximum planet--to--star flux contrast $\epsilon_{max}$ along the orbital motion and defines the scattering properties of the atmosphere.
\end{itemize}

\subsection{Properties of the Cross-Correlation Function}

\citet{Charbonneau1999} and \citet{Collier1999} estimated that, even in the most favourable cases of HJs, the flux contrast in the optical domain is lower than 10$^{-4}$. This was later confirmed by, e.g., \citet{Cowan2011}, who show that the albedo of HJs ranges between 0.05 and 0.4. For example, it would take a planetary radius of $R_{\rm p}=1.7~R_{\rm J}$ and a favourable albedo $A_g=0.40$ to make 51~Peg~b shine 10$^{-4}$ times as bright as its parent star. Because these are very optimistic conditions, this means that most likely the planetary imprint in the stellar spectrum is buried inside the noise of the spectra, and it is thus out of reach even for the best current spectroscopic facilities.

The cross-correlation Function (CCF) technique has proved to be a powerful tool to boost the planetary signal and make it larger than the spectral noise \citep[e.g.][]{Snellen2010,Brogi2012}. It basically looks for the best match between an observed spectrum and a conveniently Doppler-shifted reference template, may it be a binary mask or a model spectrum. In other words, the CCF is essentially the convolution of the observed spectrum and the template in the radial-velocity space. The result of the convolution, called CCF itself, is a good approximation of the average line profile and its signal-to-noise ratio (S/N) is approximately equal to the spectral S/N multiplied by the square root of the number of absorption lines in the reference template. For the set of observed spectra we are going to analyze in this work, which have S/N$\approx$200 (Table~\ref{tab:log}), the S/N of the CCF would increase to 14,000 if the template contains 5,000 lines, typical for binary masks used to process HARPS spectra. If we also consider that in each night of observations there are at list 40 spectra, the S/N of the cumulated signal would be $\gtrsim$90,000, making it possible to detect a planetary signal as weak as $3\times10^{-5}$ with a 3$\sigma$ significance.

The aim of the CCF technique is to compute the average spectral line profile, while no emphasis is put on the spectral continuum. For this reason, the observed spectra are usually normalized to continuum, a procedure which does not affect the shape of the spectral lines. The normalization aims at avoiding any bias introduced by the shape of the continuum, and makes it possible to compare spectra taken at different epochs with different airmasses and/or weather conditions. Hereon, we will implicitly assume that the observed spectra and the model spectrum are normalized to continuum. Moreover, in the following we will not cross-correlate the individual normalized spectra $f$ and the corresponding model spectrum $f_m$, but the functions $1-f$ and $1-f_m$. In this way, the continua of the observed and model spectra are set to zero and the absorption lines are turned upside-down. The final effect is that the computation of the integral near the absorption lines provide a positive quantity, while it provides a null contribution elsewhere.

\citet{Borra2018} show the technical advantages of computing the CCF using a stellar template derived by averaging the observed spectra\footnote{\citet{Borra2018} use the nomenclature \lq\lq Auto-Correlation Function (ACF)\rq\rq\ in their work. Strictly speaking, the ACF is the cross-correlation of a signal with a copy of itself. What they actually compute is, though, the cross-correlation of the spectra with a stellar template, obtained as the average of a list of spectra. This is why we prefer to keep the wording \lq\lq CCF\rq\rq\ in the rest of our work.}. One of the most important is that the use of a binary mask may lead to mismatches in the positions and/or depths of the spectral lines, leading to the amplification of the noise in the CCF, while the average spectrum ensures a better match between spectra and templates. Secondly, the computation is less sensitive to numerical inaccuracies in the interpolation and integration processes. Noteworthy is the fact that, if a planetary signal is present, it will show up in correspondence of the radial velocity of the planet in the stellar rest frame straight away, and no correction with respect to the stellar radial velocity is needed. For these reasons we follow the approach of \citet{Borra2018}, i.e. the computation of the CCF using an average stellar spectrum, to homogeneously analyze the sets of spectra listed in Table~\ref{tab:log}.

From a theoretical point of view, if we convolve a spectrum with Gaussian shaped lines, all with the same variance $\sigma_o^2$, with a model spectrum whose lines have width $\sigma_m^2$, then the CCF is a Gaussian function with variance given by $\sigma^2=\sigma_o^2+\sigma_m^2$. In particular, if the spectrum and the model are characterized by the same $\sigma_o$, then the variance of the CCF is simply $\sigma^2=2\sigma_o^2$. We can thus model the CCF$_o$ of a stellar spectrum $f_*(v=0)$ in its rest frame and the model spectrum $f_m$ as:
\begin{equation}
CCF_o(v)=(1-f_*)\times (1-f_m) =\delta+A e^{-\frac{v^2}{2\sigma^2}},\label{eq:onlystar}
\end{equation}
where $A$ is the amplitude of the Gaussian function and $\delta$ is an offset term. The latter is due to the fact that there is some random overlap among the lines pattern in the observed and model spectra respectively, even when the two are not aligned. This means that even in case of misalignment the convolution does not return a null result. This offset $\delta$ is, in principle, a function of $v$ as it depends on how the line pattern in the observed and model spectra cross-correlate in the velocity space. As shown in Fig.~\ref{fig:exampleCCF}, departures from a constant value show up as correlated noise in the CCF continuum, whose degree of correlation depends on the line broadening in the model and observed spectra.

\subsection{The planet CCF}

Let us now assume that the observed spectrum $F$ in the stellar rest frame is the combination of the stellar spectrum $F_*$ and the spectrum reflected by the planet $F_p$ as in Eq.~\ref{eq:fp}: 
\begin{equation}
    F=F_*+F_p=F_*(v=0)+\epsilon F_*(v=v_p).\label{eq:fluxsum}
\end{equation}

The stellar spectrum can be factorized into the continuum spectrum $F_c$ and the line spectrum $f_*$, where the latter equals to 1 where there is no line absorption and decreases towards zero according to the opacity profile of the absorption lines. The $f_*$ factor thus corresponds to the normalized spectrum introduced in the previous section:
\begin{equation}
    F_*(v)=F_c(v)\cdot f_*(v),\label{eq:fluxproduct}
\end{equation}
where we have made explicit the dependency on the velocity $v$ of source. Since both $F_c$ and $f_*$ depend on $v$, then the planetary spectrum $F_p$ in principle shifts with respect to the stellar spectrum both in terms of continuum spectrum and line spectrum. 

In the general case of a planet orbiting its host star, the rotational velocity is such that the Doppler shift in the optical domain correspond to a few \AA. In the specific case of 51~Peg~b, assuming the orbital speed of 132~km/s \citep{Brogi2013,Martins2015,Birkby2017,Borra2018}, the Doppler shift at 5000~\AA\ is 2.2~\AA. We assume that this shift is not large enough to introduce a significant displacement of the continuum spectrum. In other words, we can drop the dependency of $F_c$ on $v$. Plugging Eq.~\ref{eq:fluxproduct} in Eq.~\ref{eq:fluxsum} we thus derive:
\begin{align}
    F=&F_c\cdot f_*(v=0)+\epsilon F_c\cdot f_*(v=v_p)=\nonumber\\
    =&F_c( f_*(v=0)+\epsilon \cdot f_*(v=v_p))\nonumber\\
    =&(1+\epsilon)F_c\frac{f_*(v=0)+\epsilon \cdot f_*(v=v_p)}{1+\epsilon}.\label{eq:totalspectrum}
\end{align}

The last term in Eq.~\ref{eq:totalspectrum}
\begin{equation}
f\equiv\frac{f_*(v=0)+\epsilon \cdot f_*(v=v_p)}{1+\epsilon}    
\end{equation}
has the physical meaning of a normalized spectrum, as it is equal to 1 for the wavelengths not affected by line absorption, and decrease towards zero in correspondence of the spectral lines in the stellar and planetary spectra ($f_*(v=0)$ and $f_*(v=v_p)$ respectively). In this regard, the term $(1+\epsilon)F_c$ in Eq.~\ref{eq:totalspectrum} corresponds to the continuum spectrum.

The expected contrast $\epsilon$ is of the order of $10^{-4}$ or below. We can thus compute the Taylor expansion of Eq.~\ref{eq:totalspectrum} in powers of $\epsilon$ to derive:
\begin{equation}
    f=\frac{f_*(v=0)+\epsilon f_*(v=v_p)}{1+\epsilon}\simeq(1-\epsilon)f_*(v=0)+\epsilon f_*(v=v_p).\label{eq:combined}
\end{equation}

This equation shows that the observed normalized spectrum is the weighted average of the stellar and planetary normalized spectra. Moreover, in the case of $v_p\neq0$, i.e. when the stellar and reflected spectra are not aligned, the intensity of the absorption lines in the observed spectrum is lower than the purely stellar one, as the presence of the planetary spectrum fills-in, or veils, the line component of the stellar spectrum.

Convolving the spectrum in Eq.~\ref{eq:combined} with the stellar model, and using the linearity of the convolution operator, we derive:
\begin{align}
C&CF(v)=(1-f)\times (1-f_m)=\nonumber\\
=&(1-(1-\epsilon)f_*(v=0)-\epsilon f_*(v=v_p))\times (1-f_m)=\nonumber\\
=&(1-\epsilon)(1-f_*(v=0))\times (1-f_m)+\epsilon(1-f_*(v=v_p))\times (1-f_m)=\nonumber\\
=&(1-\epsilon)CCF_o(v)+\epsilon CCF_o(v-v_p).\label{eq:ccf}
\end{align}
We hereby remark that the combined CCF is the weighted mean of the stellar and planetary CCFs.

The goal of the method is to measure the amplitude of the planetary contribution $\epsilon$ in Eq.~\ref{eq:ccf}, in order to derive the albedo $A_g$ from Eq.~\ref{eq:epsilon}. Since we expect that the contrast $\epsilon$ is of the order of 10$^{-4}$ or lower, then the expected amplitude of the planetary CCF is small and buried in the noise of the wings of the stellar CCF. This noise is difficult to quantify \textit{a priori}, as it is a mixture of a random component due to the noise in the observed spectra, and the correlated noise in the offset $\delta$ discussed above. This last term is usually the largest one at this stage. It shows the same pattern in the CCF of all the spectra and can be minimized by normalization with an average CCF profile. Again by the linearity property, the average $\overline{CCF}(v)$ can be written as:
\begin{align}
\overline{CCF}(v)=&\frac{\Sigma_i \left[(1-\epsilon_i)CCF_o(v)+\epsilon_i CCF_o(v-v_{p,i})\right]}{N}\simeq\nonumber\\
\simeq&\frac{\Sigma_i(1-\epsilon_i)}{N}CCF_o(v)+\frac{\Sigma_i \overline{\epsilon}CCF_o(v_{p,i})}{N}=\nonumber\\
=&(1-\overline{\epsilon})CCF_o(v)+\overline{\epsilon}\frac{\Sigma_i CCF_o(v_{p,i})}{N},\label{eq:CCFbar}
\end{align}
where we have assumed that the contrasts $\epsilon_i$ can be approximated by the average contrast $\overline{\epsilon}$ (this is the typical case of spectra taken within the same night of observations). Incidentally, we note that the last line in Eq.~\ref{eq:CCFbar} corresponds to the CCF of the average spectra in Eq.\ref{eq:combined}.

Plugging Eq.~\ref{eq:onlystar} in the last term of Eq.~\ref{eq:CCFbar}, after simple math we derive:
\begin{align}
\overline{CCF}(v)=&(1-\overline{\epsilon})CCF_o(v)+\overline{\epsilon}\frac{\Sigma_i\left(\delta+Ae^{-\frac{(v-v_{p,i})^2}{2\sigma^2}}\right)}{N}=\nonumber\\
=&(1-\overline{\epsilon})CCF_o(v)+\overline{\epsilon}\delta\frac{\Sigma_i1}{N}+\overline{\epsilon}A\frac{\Sigma_ie^{-\frac{(v-v_{p,i})^2}{2\sigma^2}}}{N}=\nonumber\\
=&(1-\overline{\epsilon})CCF_o(v)+\overline{\epsilon}\delta+\overline{\epsilon}A~G(v),\label{eq:CCFbar2}
\end{align}
where we define
\begin{equation}
    G(v)\equiv\frac{\Sigma_i e^{-\frac{(v-v_{p,i})^2}{2\sigma^2}}}{N}.\label{eq:G}
\end{equation}

Equation~\ref{eq:G} is the average of a set of shifted Gaussian functions and represents the dilution of the planetary signal in the average CCF depending on the velocities $v_{p,i}$ spanned by the planet. If the velocities $v_{p,i}$ differ by many $\sigma$, then the exponential terms do not overlap, and the function $G(v)$ is basically the series of $N$ Gaussian functions, each one centered at its corresponding $v_{p,i}$ and whose amplitude is $1/N$. Conversely, for our typical datasets the planetary CCFs drift by less than $\sigma$ from one observation to the next, such that the exponential functions in Eq.~\ref{eq:G} partially overlap. This means that the individual exponential contributions cannot be distinguished in the shape of the function $G(v)$, which tends to value $\lesssim$1 when $v$ runs in the velocity range encompassed by the set $v_{p,i}$, and tends to 0 as $v$ runs out of this range.

The average CCF in Eq.~\ref{eq:CCFbar2} can now be used to normalize the CCF of the individual spectra (Eq.~\ref{eq:ccf}), obtaining:
\begin{equation}
r(v)=\frac{CCF(v)}{\overline{CCF}(v)}=\frac{(1-\epsilon)CCF_o(v)+\epsilon CCF_o(v-v_p)}{(1- \overline{\epsilon})CCF_o(v)+\overline{\epsilon}\delta+\overline{\epsilon}A~G(v)}.\label{eq:ratio}
\end{equation}
$r(v)$ thus represents the amplitude of any Doppler-shifted signal with respect to the continuum of the average CCF.

\begin{figure}
    \centering
    \includegraphics[width=.9\linewidth]{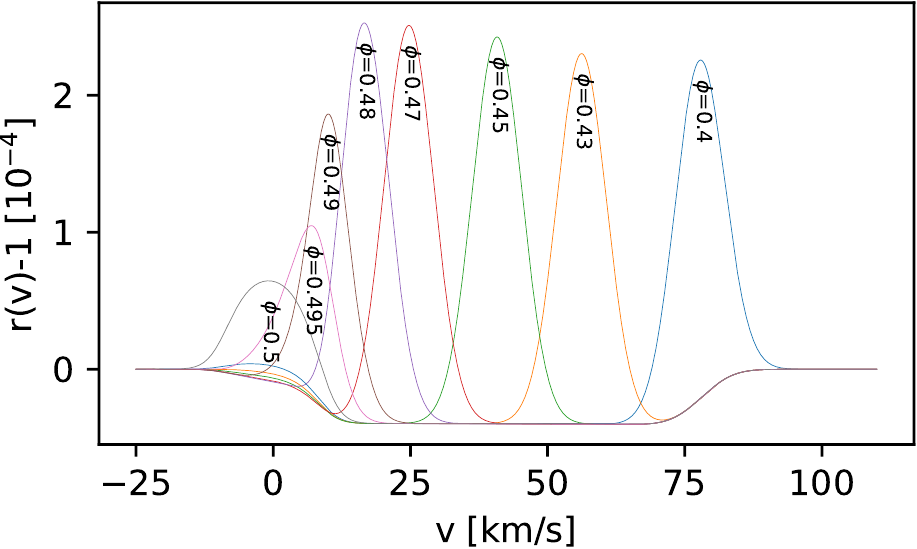}
    \caption{Expected planetary signal as discussed in the text. Different orbital phases approaching superior conjunction are simulated, as annotated in the plot. The parameters adopted for the simulations are $\epsilon_{\rm max}=10^{-4}$, $\sigma$=10.6~km/s, $A=3000$ and $\delta=1000$ (Eq.~\ref{eq:ratio}), together with the orbital solution derived in Sect.~\ref{sec:orbit}.}
    \label{fig:modelCCF}
\end{figure}

In Fig.~\ref{fig:modelCCF} we plot Eq.~\ref{eq:ratio} assuming Eq.~\ref{eq:onlystar} with $\sigma$=10.6~km/s, $A=3000$ and $\delta=1000$, which are a good approximation of the width, amplitude and continuum level respectively of the CCF of our observed spectra. We also adopt the maximum contrast of $\epsilon_{\rm max}=10^{-4}$ and the orbital inclination $i=80^\circ$ \citep{Borra2018}, together with the orbital solution obtained in Sect.~\ref{sec:orbit} to compute the phase-dependent planet--to--star flux ratio as in Eqs.~\ref{eq:phaseFunction}--\ref{eq:epsilon}. For $G(v)$ we simulate 70 observations evenly spaced in time, ranging from phase $\phi=0.4$ to $\phi=0.5$. For simplicity, we discuss three different velocity domains:

\renewcommand{\labelitemi}{\textbullet}
\begin{itemize}

\item $\mid v \mid > 4\sigma$ and $\mid v-v_p \mid > 4\sigma$:

When the stellar model is far from matching both the stellar and the planetary spectrum, all the exponential terms in Eq.~\ref{eq:ratio} are negligible, and we obtain:
\begin{equation}
    r(v)\simeq\frac{(1-\epsilon)\delta+\epsilon\delta}{(1-\overline{\epsilon})\delta+\overline{\epsilon}\delta}\simeq1.
\end{equation}
This result shows that the function $r(v)$ is indeed the normalization of the CCF. We also remark that the normalization minimizes the effects of the correlated noise pattern in the offset $\delta$.

\item $\mid v \mid < 4\sigma$ and $\mid v-v_p \mid > 4\sigma$:

When the stellar model used to compute the CCF is close to matching the stellar spectrum, and if the planetary CCF is far enough in the velocity space such not to contaminate the stellar CCF, then the exponential terms due to the planetary CCFs in Eq.~\ref{eq:ratio} are negligible. By consequence, if we approximate $\overline{\epsilon}\simeq\epsilon$, then the numerator and the denominator are the same, and we can write $r(v)\simeq1$. This result formalizes the fact that we can erase the dominant stellar signal by division with $\overline{CCF}(v)$.

\item $\mid v \mid > 4\sigma$ and $\mid v-v_p \mid < 4\sigma$:

This is the velocity range of interest that emphasizes the planetary signal while avoiding the stellar CCF. In this case, Eq.~\ref{eq:ratio} can be approximated as:
\begin{equation}
    r(v)\simeq\frac{(1-\epsilon)\delta+\epsilon\left(\delta+A e^{-\frac{(v-v_p)^2}{2\sigma^2}}\right)}{(1-\overline{\epsilon})\delta+\overline{\epsilon}\delta+\overline{\epsilon}A~G(v)}=\frac{\delta+\epsilon A e^{-\frac{(v-v_p)^2}{2\sigma^2}}}{\delta+\overline{\epsilon}A\ G(v)}
\end{equation}
In particular, the amplitude of the planetary signal is obtained substituting $v=v_p$:
\begin{equation}
    r(v_p)=\frac{\delta+\epsilon A}{\delta+\overline{\epsilon}A\ G(v_p)}.\label{eq:rvThirdCase}
\end{equation}
This result shows that the amplitude of the planetary signal is a function of the maximum contrast $\epsilon$, the stellar CCF's parameters $A$ and $\delta$, and the sampled planetary velocities $v_{p,i}$ through $G(v)$. In the best case scenario, the planetary signal in the average CCF is completely diluted such that $G(v_p)=0$ and, by consequence, the maximum signal we can extract is:
\begin{equation}
    r_{\rm max}(v_p)=\frac{\delta+\epsilon A}{\delta}=1+\epsilon\frac{A}{\delta},
\end{equation}
i.e.\ the maximum planetary signal would have the amplitude $\epsilon A/\delta$ over the continuum.

In a more realistic scenario, we can not neglect the contribution of $G(v_p)$, which being a positive quantity reduces the signal $r(v_p)$. As a matter of fact, by making the approximation $\overline{\epsilon}\simeq\epsilon$ and by means of Taylor expansion in powers of $\epsilon$, we can rewrite Eq.~\ref{eq:rvThirdCase} as:

\begin{align}
r(v_p)\simeq&\frac{\delta+\epsilon A}{\delta+\epsilon A\ G(v_p)}\simeq\nonumber\\
\simeq&1+\epsilon\frac{A}{\delta}(1-G(v_p)),
\end{align}
i.e.\ the contrast of the planetary signal against the continuum is reduced by a factor $1-G(v_p)$ with respect to $r_{\rm max}(v_p)$. For the datasets we analyze in this work, following the definition in Eq.~\ref{eq:G} we have $G(v_p)\simeq10^{-2}-10^{-1}$, i.e.\ the planetary signal is reduced by 10\% at most.

From a different perspective, the effect of $G(v_p)$ in Fig.~\ref{fig:exampleCCF} is to lower the continuum of the $r(v)$ function in the velocity range spanned by the planet, such to decrease the strength of the planetary signal.
Incidentally, we remark that this effect was ignored by \citet{Martins2013} and it may explain why they could not retrieve exactly the same signal which they injected in their simulations. Moreover, the decrease in the continuum level is present in the examples shown by \citet{Borra2018}, but the authors do not discuss its origins and effects.

\end{itemize}

Equation~\ref{eq:rvThirdCase} and Fig.~\ref{fig:modelCCF} formalize the fact that the best orbital phases to sample to maximize the amplitude of the planetary signal are those closest to superior conjunction. Most importantly, they show that when several spectra taken during the same night are averaged, the planetary signal in the average CCF is diluted over the orbital velocities. The direct effect is that, if the observations cover a conveniently large range of orbital velocities, the amplitude of the planetary signal in the average CCF is greatly reduced, and the normalization does not cancel the individual planetary CCFs. Nonetheless, we find that for $0.48\lesssim\phi\lesssim 0.52$ the planetary CCF becomes too close (less than 4$\sigma$) to the stellar counterpart, such that the approximations in Eq.~\ref{eq:rvThirdCase} do not hold anymore. In particular, when the planetary and the stellar CCFs are less than $\sim4\sigma$ apart, i.e.\ when the planetary and stellar spectra blend in the wavelength space, they tend to mimic a purely stellar spectrum. The main effect is that the stellar and planetary CCFs cannot be resolved anymore. Substituting $v=v_p$ in Eq.~\ref{eq:ratio}, and by means of Eq.~\ref{eq:onlystar}, we obtain:
\begin{align}
    r(v)=&\frac{(1-\epsilon)CCF_o(v)+\epsilon CCF_o(v)}{(1- \overline{\epsilon})CCF_o(v)+\overline{\epsilon}\delta+\overline{\epsilon}A~G(v)}=\nonumber\\
    =&\frac{CCF_o(v)}{(1- \overline{\epsilon})CCF_o(v)+\overline{\epsilon}\delta+\overline{\epsilon}Ae^{-\frac{v^2}{2\sigma^2}}-\overline{\epsilon}Ae^{-\frac{v^2}{2\sigma^2}}+\overline{\epsilon}A~G(v)}=\nonumber\\
    =&\frac{CCF_o(v)}{CCF_o(v)-\overline{\epsilon}A\left(e^{-\frac{v^2}{2\sigma^2}}-G(v)\right)}.\label{eq:r1}
\end{align}
Due to the definition in Eq.~\ref{eq:G}, the bracketed quantity in Eq.~\ref{eq:r1} is always positive, and this explains the bump in Fig.~\ref{fig:modelCCF} at phase $\phi=0.5$. We note that it is not easy to analyze this bump to extract the planetary CCF, both because of its mathematical formalization and the reduced amplitude compared with earlier (and later) orbital phases. We thus would relax the statement of \citet{Borra2018} according to which the planetary signal can be extracted also at superior conjunction.

\section{Observations and data reduction}\label{sec:observations}

We observed the 51~Peg system as part of the GAPS program for the TNG \citep[PI G.\ Micela, ][]{Covino2013} in GIARPS mode \citep{Claudi2017}, which allows simultaneous coverage of the optical and near-infrared spectral bands. In this work we only analyze the optical spectra provided by the HARPS-N instrument \citep{Cosentino2012}, a collection of two sets of observations in the nights of 26 and 27 of July 2017, for a total of 159 spectra. The infrared spectra are currently under analysis and they will be discussed in a future publication. We complement our dataset with publicly available data, taken with the same purpose of measuring the light reflected by the HJ in the system. The full list of dataset is in Table~\ref{tab:log}.

The dataset of program 091.C-0271 has been analyzed by \citet{Martins2015}, who claim a positive detection of the reflected light and quantify the flux ratio between the planet and the star of the order of $10^{-4}$. Their claim has been also confirmed by \citet{Borra2018} with an improved data analysis. \citet{DiMarcantonio2019} reanalyze the same data using the Independent Component Analysis \citep{Hyvarinen2001} and they attempt to recover the possible reflected spectrum of 51~Peg~b gave no conclusive results. Authors report that the usage of ICA methodology to extract reflected spectrum from the host star is a novel technique and simulations had shown that requirements on SNR are more stringent. Despite this, a low detection significance has been obtained even though with a different estimator if compared with the work of by \citet{Martins2015} and \citet{Borra2018}, which leads the authors to be cautious in claiming reflected light detection.

As we will discuss in the following, our data analysis requires that many spectra are taken within the same night of observation. For this reason, we do not analyze the full library of spectra from the HARPS archive, as they were sparsely collected across different nights (see the itemized list above for the dates spanned by each program). We only select the 39 (out of 91) spectra taken on the night of 2013-09-20 for program 091.C-0271 and the 48 (out of 218) spectra taken on the night 2018-10-21 for program 101.C-0106. These two subsets are the largest collections of back-to-back spectra provided by the two programs. The remaining spectra have been collected occasionally on different dates and we use them only for the refinement of the orbital solution (Sect.~\ref{sec:orbit}), not for the extraction of the reflected spectrum (Sect.~\ref{sec:analysis}). Moreover, we reject the last 31 spectra of night 2016-11-02 (program CAT16B\_43) as they were taken during bad weather conditions. All the collected spectra have been taken in proximity of the superior conjunction of the planet, such as to maximize the planetary phase function (Eq.~\ref{eq:phaseFunction}). The final number of analyzed spectra in each program is reported in Table~\ref{tab:log}.

\begin{table*}
\begin{center}
\caption{Log of the observations analyzed in this work. The orbital phases are obtained using the ephemeris computed in Sect.~\ref{sec:orbit}.}\label{tab:log}
\begin{tabular}{lllllll}
\hline\hline
Date & Program & P.I. & N. of spectra\tablefootmark{a} & Exptime (s) & SN46\tablefootmark{b} & Orbital phases\tablefootmark{c}\\
\hline
2013-09-30 & 091.C-0271 & N.C.\ Santos & 39 (91) & 450 & 150--350 & 0.396--0.445\\
2015-10-27 & CAT15B\_146 & S.\ Hoyer & 76 (76) & 200 & 200--300 & 0.518--0.564\\
2016-10-12 & CAT16B\_146 & S.\ Hoyer & 63 (63) & 200 & 200--300 & 0.474--0.526\\
2016-10-29 & CAT16B\_43  & R.\ Alonso & 59 (59) & 200 & 100-200 & 0.489--0.545\\
2016-11-02 & CAT16B\_43  & R.\ Alonso & 45 (45) & 200 & 200-300 & 0.436--0.463\\
2017-07-26 & GAPS       & G.\ Micela & 78 (78) & 200 & 150--250 & 0.359--0.407\\
2017-07-27 & GAPS       & G.\ Micela & 81 (81) & 200 & 200--300 & 0.593--0.642\\
2018-08-21 & 101.C-0106 & J.H.C.\ Martins & 48 (218) & 300 & 200--300 & 0.570--0.612\\
\hline
\end{tabular}
\tablefoot{
        \tablefoottext{a}{Number of spectra used for the extraction of the reflected spectrum in Sect.~\ref{sec:analysis}. Bracketed numbers indicate the total number of spectra used for the update of the orbital solution in Sect.~\ref{sec:orbit}.}
        \tablefoottext{b}{S/N in the 46th echelle order as computed by the data reduction pipeline}
        \tablefoottext{c}{Phase range covered by the spectra used to extract the reflected spectrum (Sect.~\ref{sec:analysis}).}
}
\end{center}
\end{table*}

Our approach for data reduction and analysis works separately for each night of observation. The following description of the workflow thus applies on a night-by-night basis. Only at the end we will merge the nightly results, in order to boost the signal detection.

For each night of observation, we analyze the s1d spectra provided by the DRS pipeline using the SLOPpy (Spectral Lines Of Planets with python) pipeline \citet{Sicilia2020}. SLOPpy is a user-friendly, standard and reliable tool that is optimized for the spectral reduction and the extraction of transmission planetary spectra obtained from high-resolution observations. To this purpose, SLOPpy first applies several data reduction steps that are required to correct the input spectra for sky emission, atmospheric dispersion and presence of telluric features and interstellar lines. These last reduction steps are not performed by the DRS pipeline. Even though our aim is not the extraction of a transmission spectrum, we use the SLOPpy pipeline as its reduction steps are designed to preserve the planetary signal.

\begin{figure}
    \centering
    \includegraphics[width=.9\linewidth]{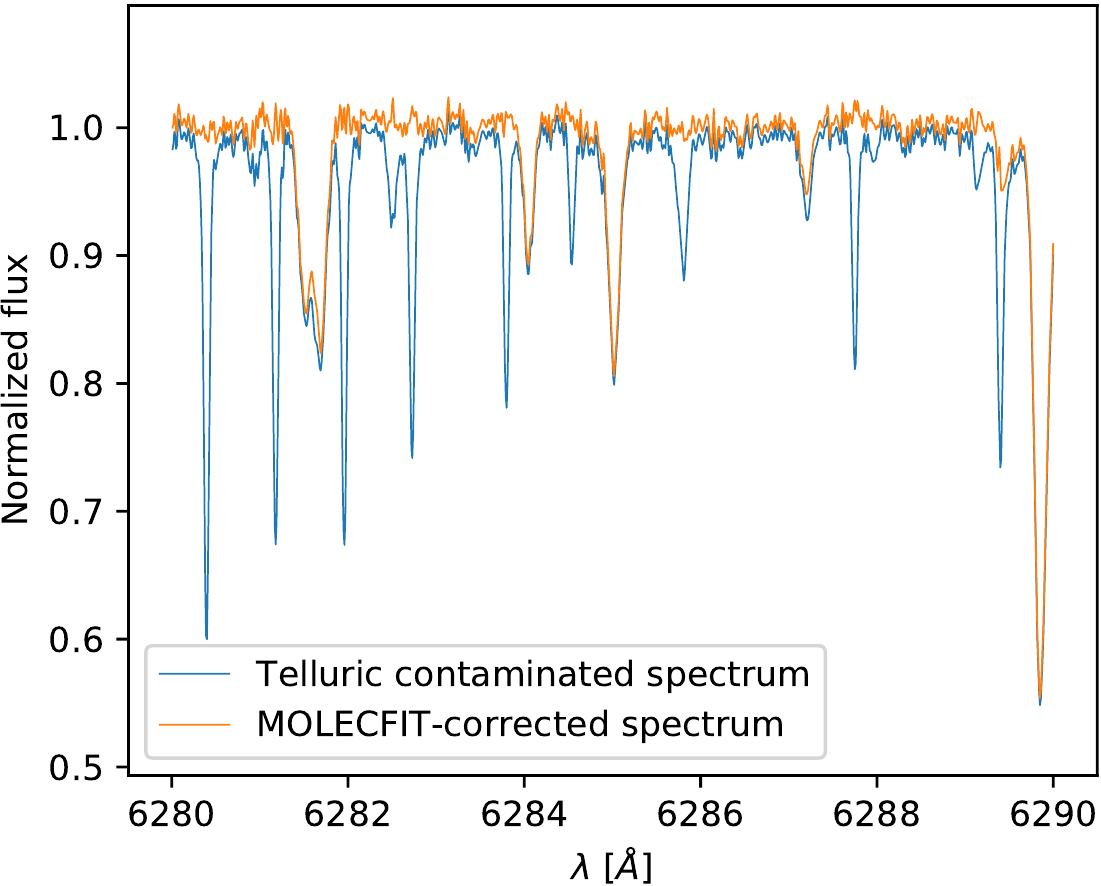}
    \caption{Example of the result of the telluric correction discussed in the text.}
    \label{fig:tellCorr}
\end{figure}

The telluric correction is performed inside SLOPpy using MOLECFIT \citep{molecfit1, molecfit2}. In order to compute the best telluric model, we consider those wavelength ranges not contaminated by stellar lines to inject to MOLECFIT. We make the selection of wavelength ranges only once per night and we use it for all the spectra of the same night. This approach is motivated by the fact that during the night the stellar and telluric spectra do not shift significantly with respect to each other, and blends involve the same group of lines along the series of spectra. By visually checking the result of the telluric removal, we find that no residuals are left above the noise level, with the exception of some left-overs comparable with spectral noise for the O$_2$ lines at wavelengths longer than $\sim$6250~\AA~(Fig.~\ref{fig:tellCorr}). In Sect.~\ref{sec:analysis} we will check that these systematic residuals do not hamper the CCF analysis.

We set up SLOPpy such that the reduced spectra are shifted in the stellar rest frame using the RV measured by the data reduction pipeline and interpolated over the same wavelength grid. 

Due to differential refraction, the average continuum level of the spectra can show a flux imbalance as a function of the airmass which, if not corrected, may affect the telluric correction and the whole analysis. The SLOPpy pipeline models this effect using a low-order polynomial or a spline, depending on the cases, and recalibrates the spectra to the same continuum using this model. We use these recalibrated spectra to remove some spikes, likely due to cosmic rays hits. For each wavelength bin, we compute the median and the Median Absolute Deviation (MAD) of the fluxes, we reject all the pixel values which deviate more than 5 MAD from the median and substitute them with the median flux. This typically corrects only a few pixels, or small groups of pixels, per spectrum.

After the removal of spikes, we refine the alignment of the spectra. To do so, we select the spectrum with the best S/N in the series and align all the remaining spectra by maximizing the cross-correlation with the selected high S/N spectrum. The most important aspect here is that the best alignment among the spectra is ensured, while the absolute radial velocity calibration, which is now the same for all the spectra, does not bias the search for the planetary signal, as shown in Sect.~\ref{sec:background}.

In Sect.~\ref{sec:background} we also explain why it is convenient to work with normalized spectra. We perform spectral normalization in the following way:
\begin{itemize}
\item we mask out the wavelength ranges 4815--4845~\AA, 5130--5210~\AA, 5887.5--5897.5~\AA, 6552.8--6572.8~\AA, which contain the broad H$\rm\beta$ line, \ion{Mg}{i} triplet, \ion{Na}{i} doublet and H$\rm\alpha$ line respectively;
\item we divide the spectra in 50 bins with the same width, and for each bin we compute the median value after clipping the absorption lines;
\item we interpolate the 50 median values over the original wavelength grid using a spline function, thus obtaining the continuum spectrum used for normalization purposes.
\end{itemize}

We remark here that neither the spike removal nor the normalization are expected to interfere with the planetary signal, if present. As a matter of fact, the former acts sparsely on a few pixels and only in some spectra of the series, while the latter operates on wavelength scales much wider than the FWHM of the spectral lines.

Finally, once the data reduction is complete, we first compute the reference spectrum of each night of observation as the median-average of the series of spectra, and then compute the residuals of each observed spectrum with respect to the corresponding reference spectrum. This residual spectrum is thus processed through a moving average algorithm to extract the noise model. This procedure is done individually for each spectrum as the noise model may vary with time according, e.g., to airmass or changing weather conditions. The noise model, one for each spectrum, will be useful in Sect.~\ref{sec:analysis} where we will test our analysis algorithm. The noise model we compute is consistent with the noise estimated by the HARPS and HARPS-N data reduction pipelines, and does not show the typical artifacts which occur when the spectra are not perfectly aligned with the template.

\section{Orbital solution}\label{sec:orbit}

To refine the ephemeris of 51~Peg~b, we use the list of RV data already piled-up by \citet{Birkby2017}, which consists in 639 measurements by several instruments (ELODIE, Lick, HIRES, HARPS) running from BJD=2\,449\,611 (September 1994) to BJD=2\,456\,847 (July 2014). The HARPS RV measurements in this collection correspond to the program 091.C-0271 analyzed in this work (see Table~\ref{tab:log}). Since we noticed slight differences in the times of observations and RV uncertainties with what is provided by the HARPS data reduction pipeline, for consistency we update the collection of \citet{Birkby2017}. Finally, we update and extend the same collection with the more recent programs listed in Table~\ref{tab:log}. The final list contains 1260 RV measurements (Table~\ref{tab:rvs}).

\begin{table}
\begin{center}
\caption{List of RV measurements analyzed for the refinement of the orbital solution.}\label{tab:rvs}
\begin{tabular}{cccc}
\hline\hline
 BJD$_{\rm TDB}$-2\,400\,000 & RV [m/s] & $\rm\sigma_{RV}$ [m/s] & instrument\\
\hline
  49610.53275500   &   -33258.0000  &    9.000  & ELODIE\\
  49612.47165600   &   -33225.0000  &    9.000  & ELODIE\\
  49655.31126300   &   -33272.0000  &    7.000  & ELODIE\\
\dots & \dots & \dots & \dots\\
\hline
\end{tabular}
\tablefoot{
        The complete table is made available in electronic form at the CDS.
}
\end{center}
\end{table}

We fit the RV measurements using the PyORBIT package\footnote{\url{https://github.com/LucaMalavolta/PyORBIT}} \citep{Malavolta2016}, trying both the circular and the eccentric keplerian models. The eccentric fit resulted in a negligible eccentricity ($e=0.007\pm0.003$) according to the Lucy \& Sweeney criterion \citep{Lucy1971}, consistently with previous analysis \citep[e.g.\ ][]{Naef2004, Birkby2017}. Moreover, we find no significant change in the other orbital parameters between the eccentric and circular fits. We thus report the results of the fit of the circular model.

The priors on the orbital period $P$ and the RV semi-amplitude $K$ were set to be uniform and centered on the estimates already available in the literature, but much larger than the corresponding uncertainties, resulting in uninformative priors (Table~\ref{tab:rvfit}). For each instrument we also fit an independent jitter term to account for different instrumental white noise levels and under-estimation of the uncertainties by the different reduction pipelines. An independent RV offset for each instrumental setup is also included. For the HARPS@ESO data we set two independent offsets to account for the upgrade of the fiber and the possible offset drift \citep{Locurto2015}. We adopt a similar approach for the three data series from the Lick observatory, taken with different upgrades of the instrument.

Following \citet[][and references therein]{Birkby2017}, we also explore the possibility that the data contain evidence of a long-term trend, a controversial claim which has not been firmly confirmed or disproved yet. We find no evidence of such a trend, and since the orbital parameters do not change significantly if we add a linear term to the fit, in this paper we report the results assuming the simpler model with no long-term drift.

\begin{figure*}
    \centering
    \includegraphics[width=\linewidth]{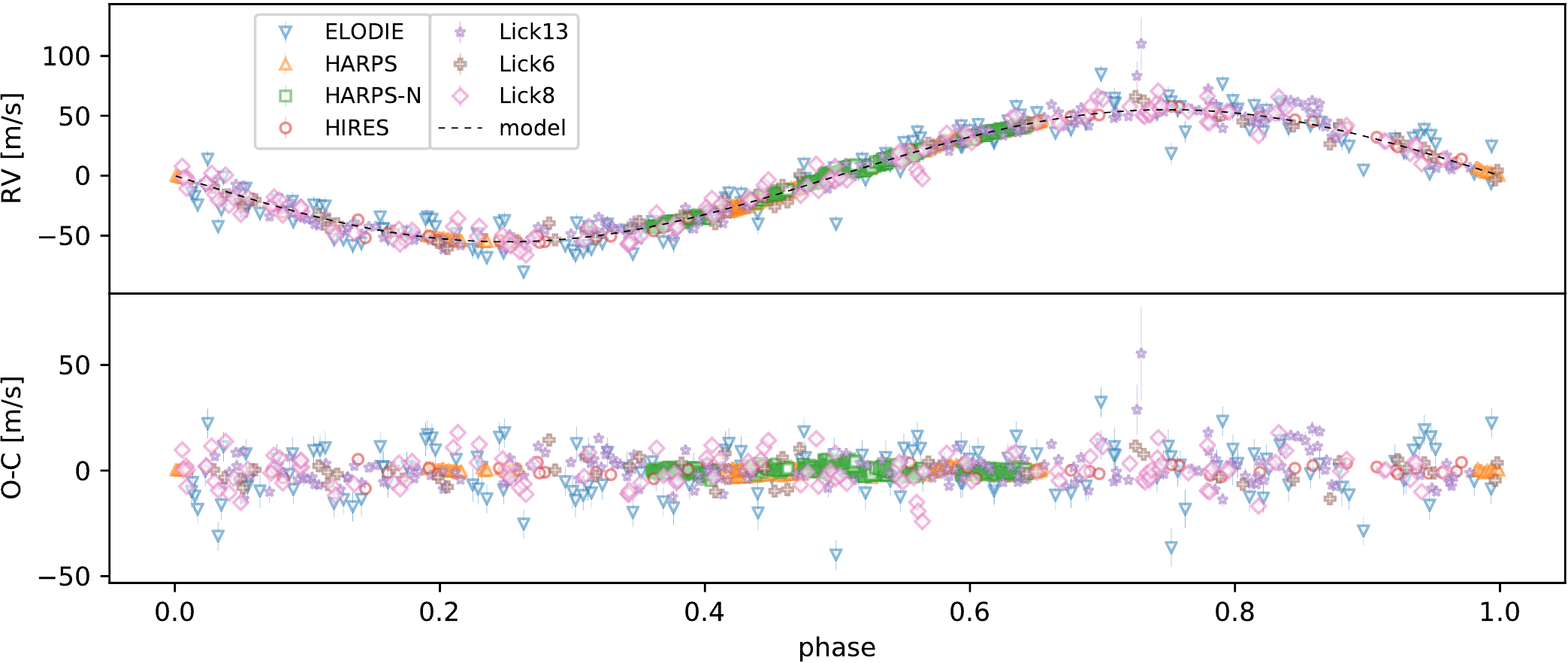}
    \caption{Phase-folded diagram of the analyzed RV measurements using the ephemeris listed in Table~\ref{tab:rvfit} (top panel) and corresponding residuals (bottom panel). Measurements from different instruments are marked with different symbols as shown in the legend. The \lq\lq Lick6\rq\rq, \lq\lq Lick8\rq\rq\ and \lq\lq Lick13\rq\rq\ labels are the same as in \citet{Birkby2017} and denote the dewar associated to the spectrograph in use during the observations. Phase 0 corresponds to the inferior conjunction of the planet.}
    \label{fig:rvfit}
\end{figure*}

\begin{figure}
    \centering
    \includegraphics[width=\linewidth]{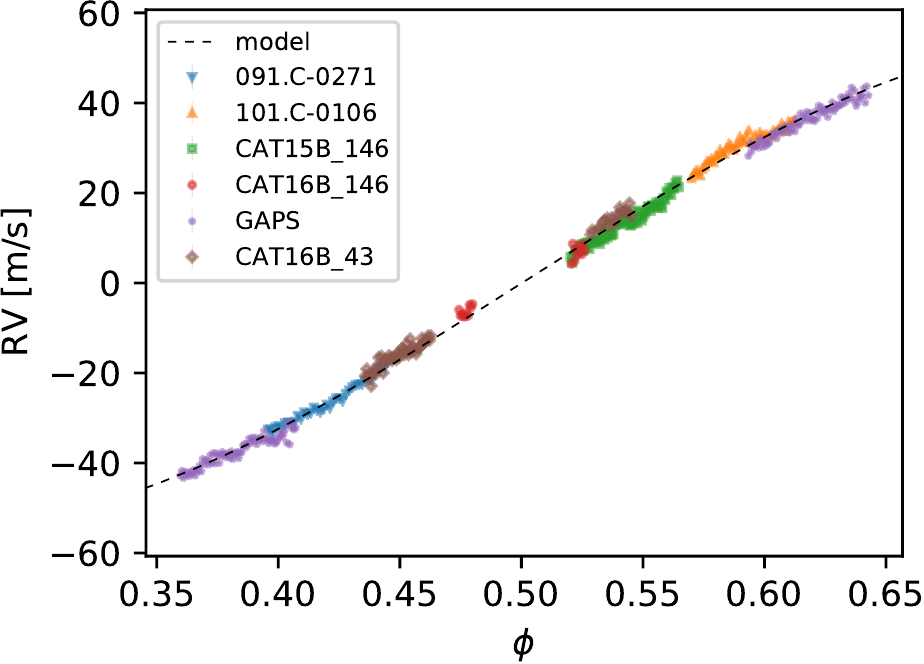}
    \caption{Phase-folded diagram of the RV measurements relative to the spectra used for the extraction of the planetary CCF (Sec.~\ref{sec:analysis}) and grouped by instrument as listed in Table~\ref{tab:rvfit}. Uncertainties are smaller than the symbol size. Phase 0 corresponds to the inferior conjunction of the planet.}
    \label{fig:rvfit_multi}
\end{figure}

We let the Monte Carlo code run for 100,000 steps, which turns out to be as long as $\sim$300 times the auto-correlation length of the chains, computed following \citet{Goodman2010}. This indicates that the fit has successfully converged, as suggested by \citet{Sokal1997} and adapted to parallel Monte Carlo chains in \url{https://dfm.io/posts/autocorr/}. Moreover, the obtained posterior distributions look nicely centered on the Maximum-A-Posteriori (MAP) best-fitting values, reported in Table~\ref{tab:rvfit} together with the corresponding 16\%--84\% quantiles. Our results are in general agreement within 2$\sigma$ with the latest ephemeris published by \citet{Birkby2017}. The best fit model and the residuals are shown in Fig.~\ref{fig:rvfit}, while Fig.~\ref{fig:rvfit_multi} shows the subset of RV measurements relative to the spectra used for the extraction of the planetary signal (Sec.~\ref{sec:analysis}).

\begin{table*}
\begin{center}
\caption{Updated orbital solution for 51~Peg~b. The best fit values are expressed as the median of the posterior distributions and the corresponding 16\%--84\% quantiles.}\label{tab:rvfit}
\begin{tabular}{llll}
\hline\hline
 & Prior & Value & Units\\
\hline
Fitted parameter: & & & \\
$P$ & $\mathscr{U}(4.2303,4.2313)$ & $4.230784\pm4\cdot 10^{-6}$ & Period (days)\\
$K$ & $\mathscr{U}(35,75)$ & $55.2\pm0.1$ & RV semi-amplitude (m/s)\\
$\phi$ & $\mathscr{U}(1.264335,1.264335+2\pi)$ & $4.406\pm0.003$ & mean longitude (rad)\\
\hline
Derived parameter: & & & \\
$T_{\rm c}$ & & $2458002.322\pm0.002$ & Time of inferior conjunction (BJD$_{TDB}$)\\
$a$ & & $0.0524\pm0.0005$\tablefootmark{a} & Semi-major axis (AU)\\
$M_{\rm p}\sin i$ & & $0.459\pm0.009$\tablefootmark{a} & planetary minimum mass (M$_{Jup}$)\\
\hline
\end{tabular}
\tablefoot{
        \tablefoottext{a}{Estimate obtained using the Gaussian prior on the stellar mass $\mathscr{N}(1.07,0.03)$.}
}
\end{center}
\end{table*}

The prior on stellar mass $M_*=1.07\pm0.03M_\sun$ used to compute the planetary mass is obtained using the PARAM web interface version 1.5 \citep[\url{http://stev.oapd.inaf.it/cgi-bin/param}, ][]{2006A&A...458..609D,2014MNRAS.445.2758R,2017MNRAS.467.1433R}, with the spectroscopic parameters $T_{\rm eff}=5814\pm19$~K, [Fe/H]=0.21$\pm$0.01 dex and $\log g =4.35\pm0.03$~dex \citep{Santos2018} as listed in the SWEET-Cat catalog \citep[https://www.astro.up.pt/resources/sweet-cat/, ][]{2013A&A...556A.150S}. The parallax $\overline{\omega}=64.65\pm0.12$~mas is taken from the Gaia DR2 \citep{Gaia2018}, while the near-infrared magnitudes are taken from the 2MASS catalog \citep{Skrutskie2006}. The stellar luminosity and the asteroseismic parameters are left undefined in PARAM, and default options are used for the computation. The uncertainty on the stellar mass takes into account the difference between the independent estimates provided by PARAM when using the two different sets of implemented evolutionary models (PARSEC \citep{Bressan2012} and MESA \citep{2017MNRAS.467.1433R} isochrones).

\section{Data analysis}\label{sec:analysis}
To enhance the detectability of the stellar light reflected by 51~Peg~b, we use the CCF technique described in Sect.~\ref{sec:background}. In the computation of the CCF, we do not use the full wavelength coverage of the spectra ($\sim$3800--6900~\AA), but we operate the following cuts referred to the stellar rest frame:
\begin{itemize}
    \item we discard the spectral range $\lambda$<4500~\AA, because the bluest echelle orders are the noisiest ones and accurate continuum normalization cannot be achieved;
    \item we discard the range $\lambda$>6700~\AA\ as it is heavily contaminated by saturated telluric absorption by O$_{\rm 2}$ that cannot be corrected accurately by MOLECFIT;
    \item the HARPS spectra do not cover a wavelength window of $\sim$100~\AA\ around 5300~\AA. To make the datasets comparable with each other, we cut the 5250--5350~\AA\ wavelength range in all the spectra in Table~\ref{tab:log}, as it is the shortest cut which excludes the blind range in all HARPS spectra;
    \item in Sect.~\ref{sec:background} we explain why the width of the CCF increases with the width of the line profile of the model spectrum. The model spectrum we use is the median-average of the observed spectra, which contain, among the others, a variety of broad lines. This has two main effects which we want to avoid: the increase of both the width and the correlated noise of the CCF. For these reasons, we cut the spectral ranges containing all the lines that, after a by-eye inspection, clearly show broadened Lorentzian profiles. These ranges are listed in Table~\ref{tab:excluded}.
\end{itemize}

\begin{table}
\begin{center}
\caption{Spectral ranges excluded in the computation of the CCF.}\label{tab:excluded}
\begin{tabular}{cl}
\hline\hline
 Spectral range (\AA) & Motivation\\
\hline
4854--4870 & H$\rm\beta$\\
4890--4893 & strong \ion{Fe}{i} lines\\
4918--4922 & strong \ion{Fe}{i} line\\
4956--4960 & strong \ion{Fe}{i} lines\\
5164--5175 & strong \ion{Fe}{i} and \ion{Mg}{i} lines\\
5181--5187 & strong \ion{Mg}{i} line\\
5887--5898 & \ion{Na}{I} D$_{\rm 1,2}$ doublet\\
6554--6574 & H$\rm\alpha$\\
\hline
\end{tabular}
\end{center}
\end{table}

\subsection{Analysis of simulated datasets}

\begin{figure}
    \centering
    \includegraphics[trim=0 0 181 15,clip]{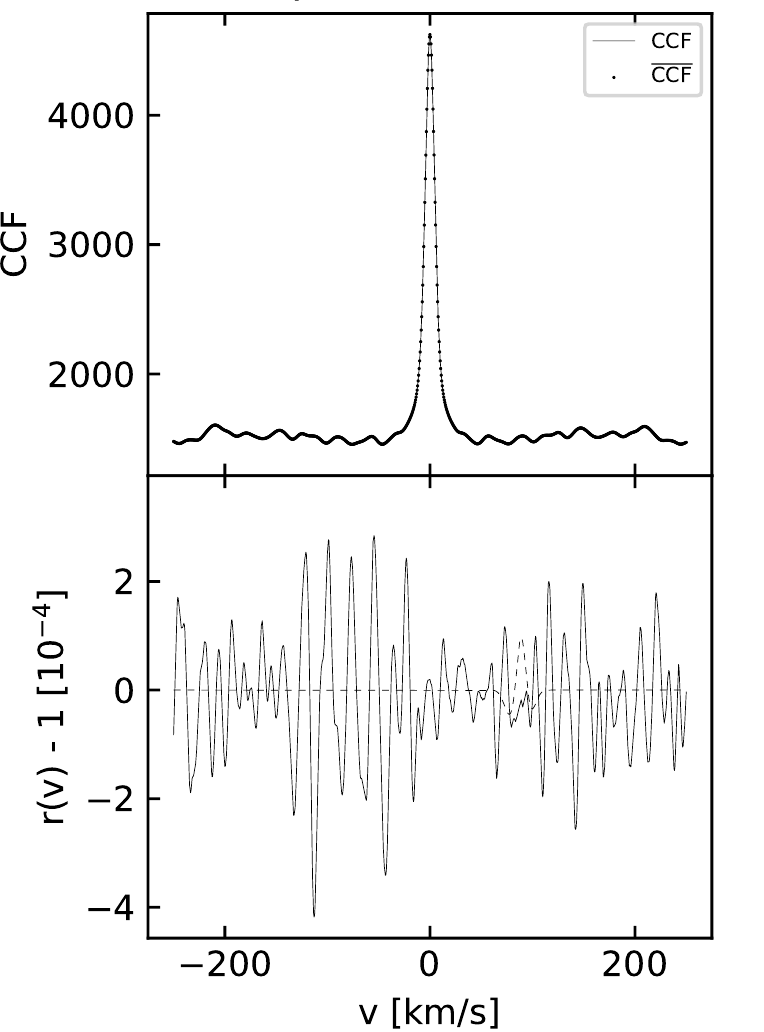}
    \caption{Example of the test discussed in the text. \textit{Top panel - }Comparison between the individual CCF and the average CCF of the simulated stellar spectra. \textit{Bottom panel - }Normalization of the CCF shown in the top panel. The dashed smooth line shows the expected noiseless signal as in Eq.~\ref{eq:rvThirdCase}, thus marking the position of the injected planetary CCF.}
    \label{fig:exampleCCF}
\end{figure}

Before analyzing the data, we run a few simulations to test the robustness of the method. The first step is thus to simulate datasets where we inject a known signal. To this purpose, for each night of observations, we use the corresponding average spectrum as a model template and we generate the simulated spectra assuming the same orbital phases sampled during the night. We inject the planetary spectrum assuming the ephemeris in Sect.~\ref{sec:orbit}, together with $K_p$=132~km/s, $\epsilon_{max}=10^{-4}$ and $i=80^\circ$ \citep[consistently with ][]{Borra2018}, and computing the planetary spectrum, velocities and phase functions according to Eqs.~\ref{eq:fp}--\ref{eq:epsilon}. To each simulated spectrum, we finally add random noise using the noise model we mention in Sect.~\ref{sec:observations}. In Fig.~\ref{fig:exampleCCF} we plot an example of the simulation, using the dataset of date 2017-07-27 in Table~\ref{tab:log}: the individual CCF and the average $\overline{CCF}$ are identical within noise, such that the planetary signal cannot be discerned. Even after normalization, the noise in the $r(v)$ function is of the order of a few 10$^{-4}$, thus comparable with the injected signal. Hence, the analysis of the individual CCFs cannot lead to the detection of the expected signal.

To enhance our capability to detect the planetary signal, we adopt the approach of \citet{Martins2015} and \citet{Borra2018}. In principle, we do not know where the planetary CCF is located with respect to the stellar CCF, as we do not know in advance the value of $K_p$ to plug in Eq.~\ref{eq:vp}. We thus build a grid of tentative $K_p$ values and, for each one, we compute the radial velocities corresponding to the phases sampled by the observations. For each tentative $K_p$ value we can thus re-center all the $r(v)$ functions in the corresponding planetary reference frame such that: if the assumed $K_p$ were correct, the planetary signals would thus be all centered at $v=0$. The main assumption of this procedure is that the average amplitude $\overline{r}(K_p)$ is maximum when the correct value of $K_p$ is used to re-center the CCFs. Conversely, if the assumed $K_p$ is wrong, then the planetary signals of the wrongly re-centered $r(v)$ functions do not match in the velocity space. In all these cases, $\overline{r}(K_p)$ is expected to be distributed around 0 with a standard deviation approximately given by the noise of the original $r(v)$ scaled down by a factor $\sqrt{N}$, where $N$ is the number of spectra.

We further optimize this procedure by adopting two additional criteria. Firstly, we remark that during each night of observation, the flux contrast changes according to the orbital phase (Eqs.~\ref{eq:cosalpha}--\ref{eq:epsilon}). With the aim of giving more emphasis to the observations closer to superior conjunction, we compute $\overline{r}(K_p)$ weighting the set of $r(v)$ functions by the corresponding $g(\alpha)$ (Eq.~\ref{eq:phaseFunction}). This procedure will be particularly useful when we will analyze jointly all the CCFs, whose phase function $g(\alpha)$ ranges from $\sim$0.66 to $\sim$0.98. We remark here that the phase function in Eq.~\ref{eq:phaseFunction} only applies to a Lambertian spherical surface, which may not be the case of 51~Peg~b. However, in the general case of a back-scattering atmosphere any alternative to Eq.~\ref{eq:phaseFunction} is a function which monotonically increases towards superior conjunction. Using different formulations of the phase function will still put more emphasis on the spectra taken closer to superior conjunction, and will introduce second order corrections to the final result. Secondly, we exclude all the spectra taken too close to superior conjunction. Referring to Fig.~\ref{fig:modelCCF}, we exclude the range $0.48<\phi<0.52$, because in this phase interval the planetary and stellar CCFs are blended and thus cannot be separated. This rejection criterion excludes 5, 43 and 40 spectra taken on 2015-10-27, 2016-10-12 and 2016-10-29 respectively (Table~\ref{tab:log}).

\begin{figure}
    \centering
    \includegraphics[width=\linewidth]{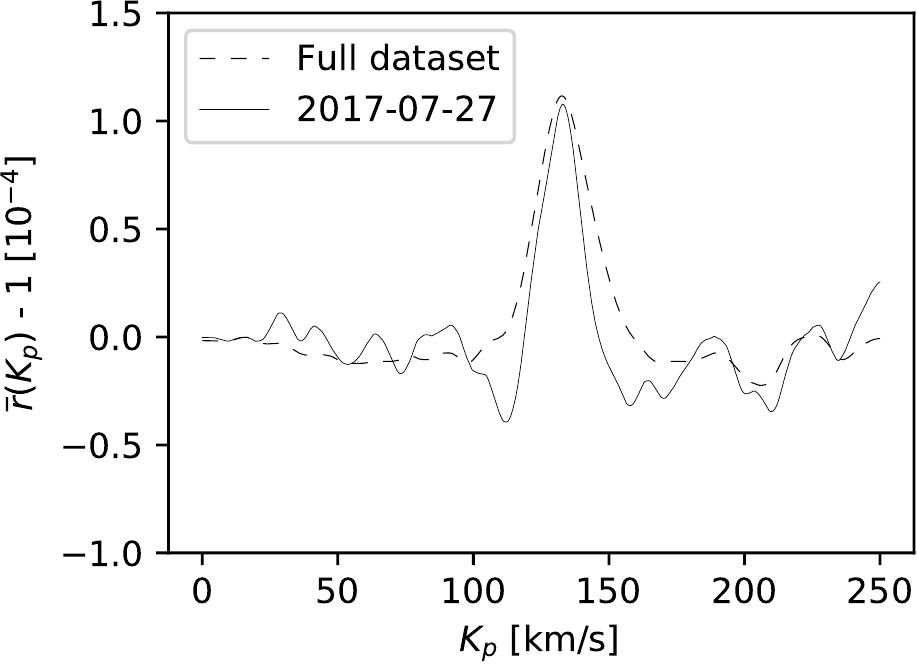}
    \caption{Simulated $\overline{r}(K_p)$ for night 2017-07-27 (solid line) and the full dataset (dotted line). The injected signal corresponds to a planet--to--star flux ratio of $\epsilon_{max}=10^{-4}$.}
    \label{fig:examplerAv}
\end{figure}

As an example, we apply this approach to the same simulated datasets discussed so far (date 2017-07-27). Figure~\ref{fig:examplerAv} clearly shows that the planetary signal peaks close to the assumed $K_p$=132~km/s. As expected, we also find that random noise in the continuum of $\overline{r}(K_p)$ has scaled down approximately by the square root of the number of spectra. We further improve the detection of the planetary signal by extending our approach to the full dataset: after exclusion of the 88 spectra close to superior conjunction, 411 spectra are left in total and random noise is further decreased by a factor of $\sim$2.

\subsection{Analysis of real datasets}

\begin{figure*}
    \centering
    \includegraphics[width=.4\linewidth]{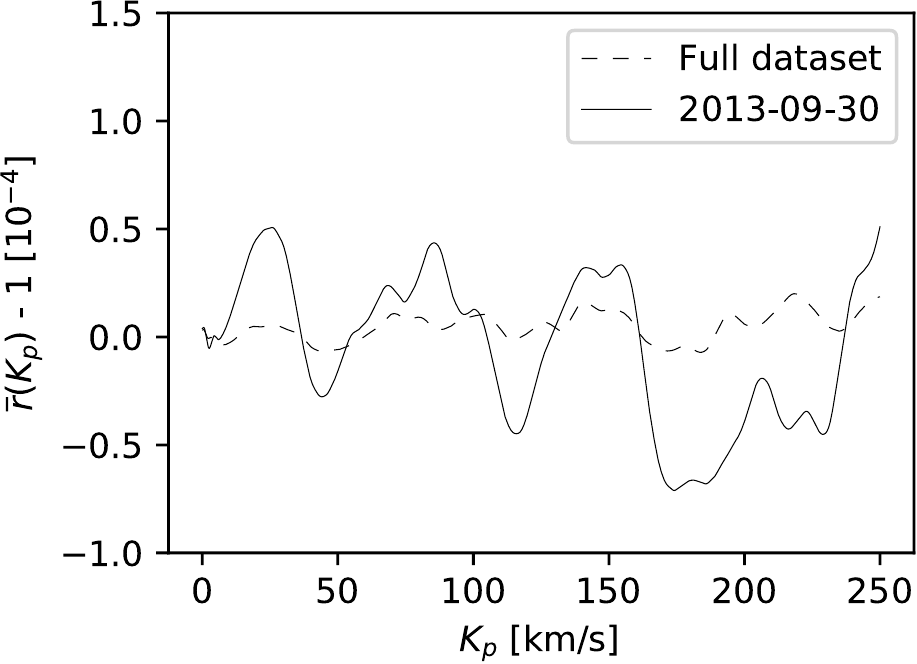}
    \hspace{2cm}
    \includegraphics[width=.4\linewidth]{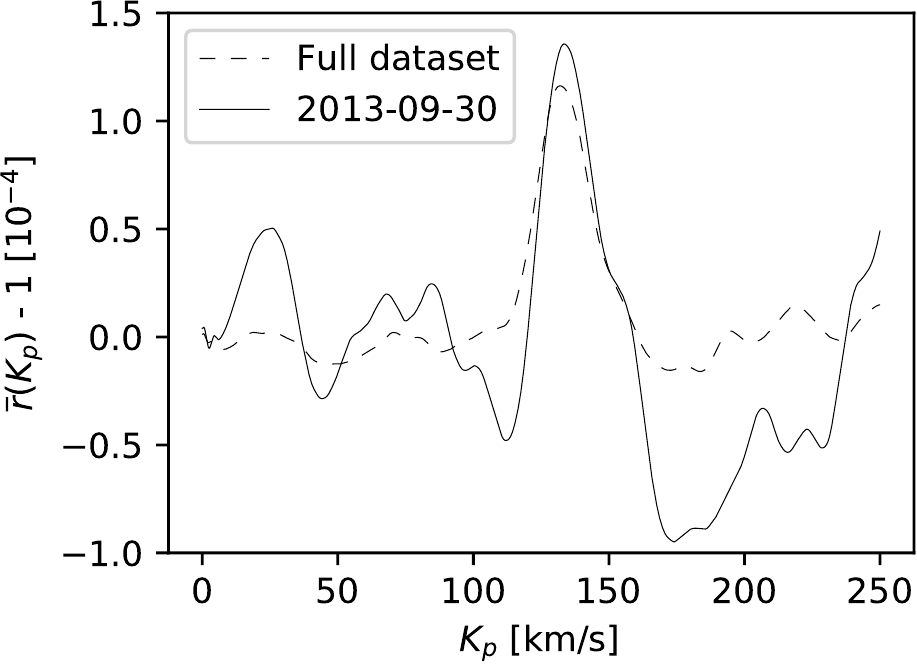}
    \caption{\textit{Left panel -- }$\overline{r}(K_p)$ for the full dataset (dashed line) and the night 2013-09-30 (solid line). \textit{Right panel -- }Same as the left panel, after the injection of a fake planetary signal with $K_p=132$~km/s and $\epsilon_{max}=10^{-4}$.}
    \label{fig:observedrAv}
\end{figure*}

The simulation discussed so far proves that our method is able to robustly extract the planetary signal claimed by \citet{Martins2015} and \citet{Borra2018}. We now want to replicate their detection using the large set of spectra we have collected (Table~\ref{tab:log}). The result of our analysis is shown in Fig.~\ref{fig:observedrAv}. The most striking evidence is that we do not find any signal at the expected $K_p\simeq$132~km/s above noise. For the sake of comparison with \citet{Martins2015} and \citet{Borra2018}, in Fig.~\ref{fig:observedrAv} we also plot the result of our method applied only to the dataset taken on 2013-09-30. Also in this case we do not find any evident signal above noise, which is now larger because we have restricted the analysis to a smaller set of spectra.

We remark that, despite all the approaches converge in using the CCF of the spectra, several differences are in place. First of all, \citet{Martins2015} and \citet{Borra2018} exclude all the wavelength ranges affected by telluric contamination. Conversely, as discussed in Sect.~\ref{sec:observations}, we carefully correct the telluric absorption in the observed spectra, and are therefore able to extend the wavelength range to analyze. This actually has the effect of reducing the noise in the CCFs and would lead us to a more robust detection, as mentioned above. To make a closer comparison in this respect, we also performed our calculations excluding the ranges affected by telluric contamination, in order to exclude the possibility that an imperfect telluric correction on our side might reduce the planetary signal. This is not the case, as this new analysis is consistent with the previous one within noise.

Secondly, we analyze only a subset of the spectra taken for the 091.C-0271 program, because the technique works best for spectra taken within a single night of observations. The reason is that the HARPS spectrograph is not designed to allow the flux calibration of the observed spectra, nor the reduction pipeline is optimized to reduce the spectra at the 10$^{-4}$ accuracy level on the flux. This may introduce some correlated noise in the continuum of the CCF (Fig.~\ref{fig:exampleCCF}). The first word of caution is that these effects may vary from night to night, depending e.g.\ on the quality of the afternoon calibration or on the thermo-mechanical parameters of the telescope. It is thus safer from this point of view to restrict the phase-resolved spectroscopic analysis within each individual night of observation.

As a matter of fact, during a given night, the Doppler shift of the stellar spectrum is less than one pixel or, in other words, the stellar spectra have a negligible Doppler shift in pixel coordinates. This means that any kind of uncorrected feature in the observed spectra does not move in wavelength with respect to the stellar spectrum. This leads to the presence of a correlated pattern in the continuum of the CCFs which does not drift in velocity space from one observation to the other. The same pattern is then propagated in the computation of the average CCF. The normalization step (Eq.~\ref{eq:ratio}) thus guarantees the correction of the correlated noise in the continuum. If several nights of observations are combined to compute the average stellar spectrum, then the result is unpredictable as it depends on how the instrumental setup has evolved and which orbital phases have been sampled.

\begin{figure}
    \centering
    \includegraphics[width=\linewidth]{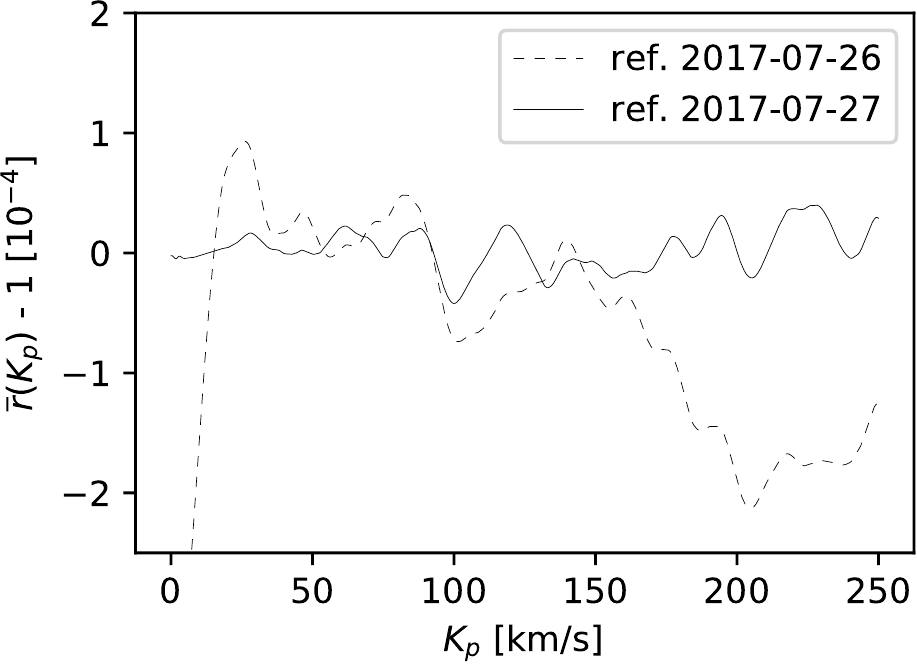}
    \caption{Comparison of the analysis of the spectra taken on date 2017-07-27 using the stellar template corresponding to the same night (solid line) and to the night of 2017-07-26.}
    \label{fig:switchNights}
\end{figure}

One possibility to minimize the systematic errors is to compare the spectra taken during two consecutive nights. In Fig.~\ref{fig:switchNights} we show the result of our analysis restricted to night 2017-07-27 using either the average spectrum of the same night or the one corresponding to night 2017-07-26. In the second case we obtain a trend which is likely due to mismatches in the spectral normalization between the two nights, while the correlated noise on top of the trend does not increase significantly.

This test also leads to another important evidence. In Sect.~\ref{sec:background} we show that the planetary signal in the average CCF is diluted and reduced in amplitude by the median-average. The dilution leads to a valley in the normalized individual CCFs at the same radial velocities of the planet in the stellar rest frame (Fig.~\ref{fig:modelCCF}). This valley partially reduces the amplitude of the planetary signal, down to the noise level in a pessimistic scenario. To maximize the planetary CCF one should thus use a reference CCF unaffected by the planetary signal in the velocity range of interest. The previous test aims at simulating such a scenario. As a matter of fact, we have analyzed the spectra of night 2017-07-27 (i.e.\ after superior conjunction) using the master spectrum of night 2017-07-26 (i.e.\ before superior conjunction). With this combination of nights, we ensure that any residual of the planetary signal in the reference CCF of night 2017-07-26 does not cover the velocity range encompassed by the expected planetary CCF on night 2017-07-27. Even in this case, where the interference of the planetary CCF with itself has been avoided, we would be able to detect the claimed planetary signal with a significance of $\sim4\sigma$, but in fact we get a negative results as shown in Fig.~\ref{fig:switchNights}.

As a final test, given that we do not find any signature of the planetary reflected spectrum, we assume that the stellar spectra are not contaminated by the planet, and we inject in the spectra a fake planetary spectrum assuming $K_p=132$~km/s and $\epsilon_{max}=10^{-4}$, as done for the simulations discussed earlier in this section. We find that the such a signal would be clearly detectable above noise in all the datasets in Table~\ref{tab:log} and even better in the joint analysis (right panel in Fig.~\ref{fig:observedrAv}). In particular, the amplitude of the planetary signal would be $\sim3.6$ and $\sim22$ times larger than noise when we analyze the night 2013-09-30 and the full dataset respectively.

\section{Discussion and conclusions}\label{sec:discussion}

The first successful detection of the planetary spectrum of 51~Peg~b is reported by \citet{Brogi2013}, who discover the absorption of carbon monoxide and water vapor in the CRIRES spectra of the dayside hemisphere. Analyzing the Doppler shift of the planetary spectra, the authors also put a constraint on the orbital inclination between 70$^\circ$.6 and 82$^\circ$.2 (with the upper limit set by the non-transiting nature of the planet) and derived the planetary velocity amplitude $K_p=134.1\pm1.8$~km/s. These measurements lead to a planetary mass of $M_{\rm p}=0.46\pm0.02~M_{\rm Jup}$. The same results are later corroborated by \citet{Birkby2017}, who also estimate the rotational velocity of the planet to be $v_{\rm rot}<5.8$~km/s.

Likewise, \citet{Martins2015} analyze the optical spectra of the 51~Peg system, looking for reflection by the planet. Using the CCF technique, they estimate the planetary velocity amplitude as $K_p=132^{+19}_{-15}$~km/s and a corresponding planetary mass of $M_p=0.46^{+0.06}_{-0.01}~M_{\rm Jup}$, thus confirming the results of \citet{Brogi2013}. The FWHM of the planetary CCF that they derived is $\sim23\pm4$~km/s, significantly broader than the stellar CCF (FWHM=7.43~km/s). The authors caution that it can be due to the fact that the signal is close to the noise level. Nonetheless, if the broadening is confirmed, according to the authors the broadening may indicate the rapid rotation of the planet (18~km/s), much faster than the tidally locked rotation (2~km/s). \citet{Strachan2020} reproduce the same broadening using a more sophisticated model which accounts for the finite size of the star and planet in the integration of radiated/scattered flux intensities across both their surfaces. \citet{Borra2018} improve the application of the CCF technique in the search of the reflected spectra, and confirm the results of \citet{Martins2015}. Moreover, they estimate a lower value for the FWHM of the planetary CCF, i.e.\ 9.69$\pm$0.28~km/s: much closer to the stellar FWHM, but still higher than predicted by \citet{Birkby2017}.

In this paper, we analyze a larger set of HARPS and HARPS-N spectra of the 51~Peg planetary system taken when the planet was near superior conjunction. In our analysis we are inspired by \citet{Martins2015} and \citet{Borra2018} in using the CCF technique as a powerful tool to extract weak signals buried in the noise. We detail the mathematical formalism about the CCF method tailored to the search of the light reflected by the planet, which we find to be only vaguely presented in the literature. We also described how we reduce the data in order to optimize the extraction of the planetary signal. We check that our method and data reduction are robust enough to allow the detection of the signal claimed by \citet{Martins2015} and \citet{Borra2018}. However, we do not find any evidence of the reflected planetary spectrum.

Including our re-analysis, there are thus two firm detections of the planetary reflected spectrum in the optical and two null detections. The two positive detections are obtained using similar techniques on the same dataset. The null detection by \citet{DiMarcantonio2019} is obtained on the same data but with a completely different mathematical approach, while our result is an extension of the CCF technique to a larger dataset. This opens the possibility that the detected signals may be caused by pitfalls of the technique coupled with the characteristics of the analyzed dataset. To this purpose, in Sect.~\ref{sec:analysis} and Fig.~\ref{fig:observedrAv} we present our algorithm run on the spectra collected on 2013-09-30, which include half the spectra analyzed by \citet{Martins2015} and \citet{Borra2018}. Even if this analysis does not return any significant signal, we notice a broad bump around $\sim150$~km/s (left panel Fig.~\ref{fig:observedrAv}). This suggests the hypothesis that the claimed detection is just a false positive signal, unluckily located at the expected velocity $K_p\sim132$~km/s, and pushed up by a wicked combination of the properties of the CCF computation with the sampled planetary orbital phases.

The controversial point about the detection is the amplitude of the planetary signal ($\epsilon_{max}=12.5\times10^{-5}$ and $8.6\times10^{-5}$ as derived by \citet{Martins2015} and \citet{Borra2018} respectively). As a matter of fact, once the planet--to--star flux ratio $\epsilon_{max}$ in Eq.~\ref{eq:epsilon} is fixed, there is an inverse proportionality between the geometric albedo $A_g$ and the square of the planetary radius $R_{\rm p}$. Based on observational evidence, \citet{Angerhausen2015} provide a typical value of $A_g=0.1$ for the planetary albedo, which translates in a radius of $\sim3.9~R_{\rm Jup}$. Combining this predicted radius with the mass estimates provided by \citet{Brogi2013} and \citet{Birkby2017}, we obtain that the bulk density expected for 51~Peg~b is 0.01~g/cm$^3$, which puts 51~Peg~b beyond the sample of H/He dominated extremely-low density planets \citep{Laughlin2018}. This is a plausible yet unlikely scenario: for example, to date in the Exoplanet Orbit Database \citep{2014PASP..126..827H} there is only one HJ \citep[HAT-P-65~b,][]{Hartman2016} out of 245 less dense than 0.15~g/cm$^3$, while the planets with the lowest density ever measured (0.03~g/cm$^3$) are the bloated Jupiter-size Earth-mass planets Kepler51~b and c \citep{Masuda2014}.

\begin{figure}
    \centering
    \includegraphics[width=\linewidth]{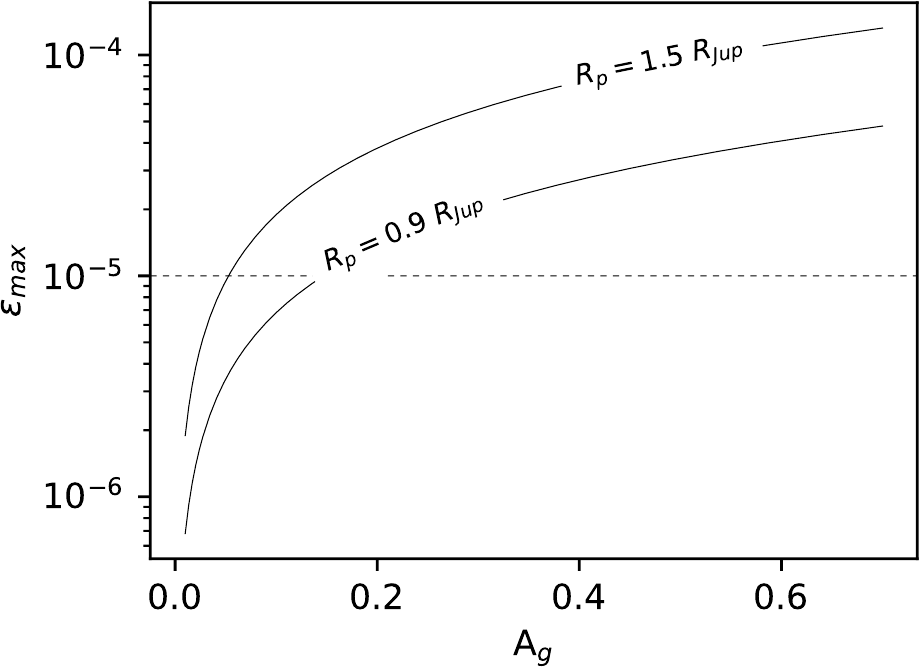}
    \caption{$\epsilon_{max}$ vs. $A_g$ relationship, as in Eq.~\ref{eq:epsilon}, assuming $R_{\rm p}=0.9~R_{\rm Jup}$ and $R_{\rm p}=1.5~R_{\rm Jup}$. The dashes represent our upper limit on $\epsilon_{max}$.}
    \label{fig:eAg}
\end{figure}

To reconcile 51~Peg~b with the general properties of HJs, one can allow higher albedos and make the corresponding planetary radii smaller. For example, following Fig.~2 in \citet{Laughlin2018} and converting densities into radii, we can assume that the maximum radius of a HJ with the same mass as 51~Peg~b is 1.5~$R_{\rm Jup}$. Inverting Eq.~\ref{eq:epsilon} and adopting the $\epsilon_{max}$ derived by \citet{Martins2015}, the corresponding planetary albedo would be $A_g\simeq0.68$. For the sake of comparison, the highest albedo reported by \citet{Angerhausen2015} is $A_g=0.32$, i.e.\ even the largest radius expected for 51~Peg~b is not able to return a feasibly low geometric albedo. 
Our analysis leads to a different scenario. Assuming that there is no trace of the planetary reflected spectrum in the data, Fig.~\ref{fig:observedrAv} (left panel) represents the noise which limits our capability to detect the planetary CCF. After some injection/retrieval experiments, we find that the minimum signal that we can detect above the 3$\sigma$ level corresponds to $\epsilon_{max}=10^{-5}$, which thus represents the upper limit for the planet--to--star flux ratio. In Fig.~\ref{fig:eAg} we compare this detection limit with the $\epsilon_{max}$ expected for 51~Peg~b. In particular, following \citet{Laughlin2018}, we assume that a HJ with the same mass as 51~Peg~b has a radius in the 0.9--1.5~$R_{\rm Jup}$ range: after plugging these radius limits in Eq.~\ref{eq:epsilon}, we plot the corresponding $\epsilon_{max}$ vs.\ $A_g$ relations in Fig.~\ref{fig:eAg}. We find that an albedo of $A_g=0.1$ corresponds to $\epsilon_{max}$ between 6$\cdot$10$^{-6}$ and 2$\cdot$10$^{-5}$, close to our detection limit. We thus conclude that our null detection is consistent with a dark ($A_g$<0.1) average-size HJ and that 51~Peg~b is not an outlier in terms of albedo and/or planetary radius. This result is consistent with the theoretical predictions provided by \citep{Sudarsky2000}: in the upper atmosphere of a HJ like 51~Peg~b the dominant contribution to the opacity is given by the broad absorption of alkali metals (\ion{Na}{}, \ion{K}{}), which precludes the silicate clouds at deeper layers from leading to a significant albedo.

\begin{acknowledgements}

This research has made use of the Exoplanet Orbit Database and the Exoplanet Data Explorer at exoplanets.org.

GSc acknowledges his niece MMa for delighting her proud uncle during the writing of this paper.

GSc, FBo, GBr, IPa and GPi acknowledge the funding support from Italian Space Agency (ASI) regulated by \lq\lq Accordo ASI-INAF n.\ 2013-016-R.0 del 9 luglio 2013 e integrazione del 9 luglio 2015\rq\rq.

GBr acknowledge support from CHEOPS ASI-INAF agreement n. 2019-29-HH.0.

MEs acknowledges the support of the DFG priority program SPP 1992 \lq\lq Exploring  the  Diversity  of Extrasolar Planets\rq\rq\ (HA  3279/12-1).

\end{acknowledgements}

\end{document}